\RequirePackage{fix-cm}
\documentclass[twocolumn]{svjour3}          
\smartqed  
\usepackage{graphicx}
\usepackage{braket}
\usepackage{siunitx} 
\usepackage{pgfplots}
\usepackage[backgroundcolor=orange, linecolor=black]{todonotes} 
\usepackage[authoryear]{natbib}
\usepackage{tikz}
\usepackage{svg}
\usetikzlibrary{shapes, arrows, positioning, decorations.pathreplacing, decorations.markings, calc}

\definecolor{blue-viridis}{rgb}{0.229739, 0.322361, 0.545706}
\definecolor{green-viridis}{rgb}{0.185783, 0.704891, 0.485273}

\tikzstyle{rectwide} = [rectangle,
                        minimum width=0.5\textwidth - 7em,
                        minimum height=1cm,
                        text centered, draw=black]
\tikzstyle{rectwidehs} = [rectangle,
                          minimum width=0.5\textwidth - 3em,
                          minimum height=1cm,
                          text centered, draw=black]
\tikzstyle{rect} = [rectangle,
                    minimum height=1cm,
                    text centered, draw=black]
\tikzstyle{rect-chiller} = [rectangle,
                    		minimum height=1.5cm,
                    	 	minimum width=1.8cm,
                    		text centered, draw=black, thick]
\tikzstyle{rect-blue} = [rectangle,
                    	 minimum height=1cm,
                    	 minimum width=1.5cm,
                    	 text centered, draw=black,
                    	 color=blue-viridis, text=black]
\tikzstyle{rect-green} = [rectangle,
                    	  minimum height=1cm,
                    	  minimum width=1.5cm,
                    	  text centered, draw=black,
                    	  color=green-viridis, text=black]
\tikzstyle{arrow2way} = [thick,<->,>=stealth]
\tikzstyle{arrow1way} = [thick,->,>=stealth]

\tikzset{pic shift/.store in=\shiftcoord, pic shift={(0,0)},
    qd/.pic={
        \begin{scope}[shift={\shiftcoord}]
        \draw[thick, fill=white] (-0.5,-0.2) rectangle(0.5,0.2);
        \draw[thick] (-0.3,0) circle[radius=0.09]; \draw[thick] (0.3,0) circle[radius=0.09];
        \draw[thick] (-0.5,0) to (-0.4,0); \draw[thick] (0.4,0) to (0.5,0);
        \draw[thick] (-0.15,0) to (0.15,0); \draw[thick] (0,0.1) to (0,-0.1);
        \draw[thick] (-0.3,0.16) to (-0.15,0) to (-0.3,-0.16);
        \draw[thick] (0.3,0.16) to (0.15,0) to (0.3,-0.16);
        
        \coordinate (-left) at (-0.5,0);
        \coordinate (-right) at (0.5,0);
        \coordinate (-anchor) at (0,0);
    \end{scope}
}}

\tikzset{pic shift/.store in=\shiftcoord, pic shift={(0,0)},
    flex/.pic={
    \begin{scope}[shift={\shiftcoord}]
        \draw ($ (-0.5,0) + (2pt,0) $) circle (2pt); \draw ($ (0.5,0) - (2pt,0) $) circle (2pt);
        \draw[thick] ($ (-0.5,0) + (2pt,0) + (-30:2pt) $) to [out=-30,in=-150] ($ (0.5,0) - (2pt,0) + (-150:2pt) $);
        \coordinate (-left) at ($ (-0.5,0)$);
        \coordinate (-right) at ($ (0.5,0)$);
        \coordinate (-anchor) at (0,0);
    \end{scope}
}}

\begin{document}

\title{A Dual-Species Atom Interferometer Payload for Operation on Sounding Rockets
}


\author{Michael Elsen \and
	Baptist Piest \and
	Fabian Adam \and
	Oliver Anton \and
	Pawe\l{} Arciszewski \and
	Wolfgang Bartosch \and
	Dennis Becker  \and
	Jonas B\"ohm \and
	S\"oren Boles \and
	Klaus D\"oringshoff \and
	Priyanka Guggilam \and
    Ortwin Hellmig \and
    Isabell Imwalle \and
	Simon Kanthak \and
    Christian K\"urbis \and
    Matthias Koch \and
	Maike Diana Lachmann \and
	Moritz Mihm \and
	Hauke M\"untinga \and
	Ayush Mani Nepal \and
	Tim Oberschulte \and
	Peter Ohr \and
	Alexandros Papakonstantinou \and
	Arnau Prat \and
	Christian Reichelt \and	
	Jan Sommer \and
	Christian Spindeldreier \and
	Marvin Warner \and
	Thijs Wendrich\and
	Andr\'e Wenzlawski\and
	Holger Blume \and
	Claus Braxmaier \and
	Daniel L\"udtke \and
	Achim Peters \and
	Ernst Maria Rasel \and
	Klaus Sengstock \and
	Andreas Wicht \and
	Patrick Windpassinger \and
    Jens Grosse	
}



\institute{
	 Michael Elsen
	  \and Marvin Warner  \and Jens Grosse \at
	Zentrum f\"ur angewandte Raumfahrttechnologie und Mikrogravitation, Universit\"at Bremen, Am Fallturm, 28359 Bremen, Germany\\
	\email{michael.elsen@zarm.uni-bremen.de}           
	\and
	Baptist Piest \and Wolfgang Bartosch \and Jonas B\"ohm \and Dennis Becker \and Matthias Koch \and Maike Diana Lachmann \and Alexandros Papakonstantinou \and Thijs Wendrich \and Ernst Maria Rasel \at
	Institut f\"ur Quantenoptik, Leibniz Universit\"at Hannover, Welfengarten 1, 30167 Hannover, Germany
	\and
	Tim Oberschulte \and Christian Spindeldreier \and Holger Blume \at
	Institut f\"ur Mikroelektronische Systeme, Leibniz Universit\"at Hannover, Appelstra{\ss}e 4, 30167 Hannover, Germany
	\and
	Oliver Anton \and Pawe\l{} Arciszewski \and Klaus D\"oringshoff \and Simon Kanthak \and Achim Peters \at
	Institut f\"ur Physik, Humboldt-Universit\"at zu Berlin, Newtonstra{\ss}e 15, 12489 Berlin, Germany
	\and
	S\"oren Boles \and Moritz Mihm \and Andr\'e Wenzlawski \and Patrick Windpassinger \at
	Institut f\"ur Physik, Johannes Gutenberg-Universit\"at Mainz, 55099 Mainz, Germany
	\and
	Fabian Adam \and Hauke M\"untinga \at 
	Institut f\"ur Satellitengeod\"asie und Inertialsensorik,  Deutsches Zentrum f\"ur Luft und Raumfahrt e.V., Callinstr. 30b, 30167 Hannover, Germany
	\and
	Claus Braxmaier \at
	Institut f\"ur Mikroelektronik, Universität Ulm, Albert-Einstein-Allee 43, 89081 Ulm, Germany\\
	Institut f\"ur Quantentechnologie, Deutsches Zentrum f\"ur Luft und Raumfahrt e.V., Wilhelm-Runge-Str. 10, 89081 Ulm, Germany
	\and
	Peter Ohr \and Ayush Nepal \and Arnau Prat \and Jan Sommer \and Daniel Lüdtke  \at
	Institut f\"ur Softwaretechnologie, Deutsches Zentrum f\"ur Luft und Raumfahrt e.V., Lilienthalplatz 7, 38108 Braunschweig, Germany
	\and
	Christian K\"urbis \and Andreas Wicht \at 
	Ferdinand-Braun-Institut, Leibniz-Institut f\"ur H\"ochstfrequenztechnik, Gustav-Kirchhoff-Str. 4, 12489 Berlin, Germany	
	\and 
	Ortwin Hellmig \and Klaus Sengstock \at
	Institute of Laser-Physics, University Hamburg, Luruper Chaussee 149, 22761 Hamburg, Germany
}

\date{Received: date / Accepted: date}

\maketitle

\begin{abstract}
\textcolor{black}{\textbf{
		We report on the design and the construction of a sounding rocket payload capable of performing atom interferometry with Bose-Einstein condensates of $^{41}$K and $^{87}$Rb.\\		
		The apparatus is designed to be launched in two consecutive missions with a VSB-30 sounding rocket and is qualified to withstand the expected vibrational loads of 1.8$\,$g root-mean-square in a frequency range between $20 - 2000\,$Hz and the expected static loads during ascent and re-entry of \SI{25}{\g}.
		We present a modular design of the scientific payload comprising a physics package, a laser system, an electronics system and a battery module. 
		A dedicated on-board software provides a largely automated process of predefined experiments.\\		
		To operate the payload safely in laboratory and flight mode, a thermal control system and ground support equipment has been implemented and will be presented.\\
		The payload presented here represents a cornerstone for future applications of matter wave interferometry with ultracold atoms on satellites.}
}
\keywords{Bose-Einstein Condensate \and Quantum Optics \and Atom Optics \and Atom
Interferometry \and Microgravity \and Sounding Rocket}
\end{abstract}

\newpage
\newpage

\section{Introduction}
\label{intro}

Space offers great potential for studies of fundamental physics with cold atoms, such as tests of the universality of free fall \citep{Aguilera2014} and the detection of gravitational waves \citep{Loriani2019}.
Utilizing free-falling atoms of two different species as sensitive probes for differential inertial forces, light-pulse atom interferometry enables to test the Einstein equivalence principle (EEP) with high accuracy \citep{Schlippert2014, Zhou2015, asenbaum2020}.
For such atom interferometers the sensitivity for measuring inertial forces acting on the atomic cloud scales quadratically with the free fall time.
One way to extend current limitations on the free fall time, typically set by the size of the baseline, is to perform measurements on microgravity platforms such as drop towers, parabola flights, sounding rockets or satellites instead of ground based setups. 
Working with long free fall times, however, requires the production of Bose-Einstein condensates (BECs) due to their small initial size and narrow momentum distribution.
The complexity of such experiments and high demands regarding stability and accuracy face us with enormous challenges when transferring the sensitive setups into space. 

Recent missions on sub-orbital and orbital platforms show a fast progress towards high precision measurements with atom optical instruments \citep{Gaaloul2022, Carollo2022, Becker2018, Liu2018,Condon2019} in microgravity environments. 
Sounding rockets offer excellent microgravity conditions and a comparably easy accessible and economically affordable platform. 
In the beginning of 2017, the sounding rocket mission MAIUS-1 (\textbf{Ma}\-te\-rie\-wel\-len\-\textbf{i}nter\-fero\-metrie \textbf{u}nter \textbf{S}chwerelosigkeit, matter wave interferometry in microgravity) successfully demonstrated the first BEC of $^{87}$Rb in space and conducted further experiments studying their coherence properties using atom interferometry \citep{Becker2018, Lachmann2021}.

To go one step further, two additional sounding rocket missions MAIUS-2/3 are planned to measure the differential accelerations of two BECs consisting of different species, $^{87}$Rb and $^{41}$K.
With these measurements, the experiments pave the way for tests of the equivalence principle in space with unprecedented sensitivity. 
The technologies developed for these missions show a pathway towards future endeaveurs on space based platforms with ultracold atoms like the international space station (ISS) \citep{Frye2019} or satellite missions \citep{Aguilera2014}. 

This paper gives a detailed overview of the payload of MAIUS-2/3, named MAIUS-B, with a focus on its technical implementation and qualification and is organized as follows: 
Section 2 gives a brief overview of the mission objectives and the experimental capabilities of the apparatus.
Section 3 will provide a description of the scientific payload which consists of different modules containing the experimental chamber, laser system and electronics system. Also, a discussion of the thermal control system (TCS) and the structure of the software is provided. In section 4, the qualification process is described and results of vibrational tests are presented.

\section{Experimental Methods}
\label{objectives}
The goal of the MAIUS-2/3 missions is to measure the differential accelerations of two free falling ultracold atom clouds of different atomic species revealing an EEP test.
Due to their difference in mass and their magnetically tunable interatomic interactions $^{87}$Rb and $^{41}$K are appropriate candidates for such studies \citep{Thalhammer2008}. Furthermore they offer optical transitions capable of laser cooling in a similar frequency range which simplifies the setup of the laser system.
Their preparation in high atom number, low expansion velocity and large overlap is crucial for the signal and the suppression of systematic effects. 
The new apparatus MAIUS-B is able to autonomously create single- or dual-species BECs containing  $^{87}$Rb and $^{41}$K with state-of-the-art repetition rates and particle numbers using an integrated atom chip.
The chip is suitable for magnetic delta-kick collimation techniques \citep{Ammann1997, Deppner2021, Corgier2020} of both species to lower their expansion velocity enabling a long pulse-separation time in the atom interferometer. 
It further allows for fast transport of the atomic ensembles to the desired interferometer region within the science chamber \citep{Corgier2018}. 
Employing optical gratings based on Raman double\-diffraction \citep{leveque, hartmann} the apparatus can perform atom interferometry with the free falling BECs. 

A typical experimental sequence to prepare these mixtures starts with loading of a three-dimensional chip magneto-optical trap (MOT, \cite{wildermuth}) with  $^{87}$Rb and $^{41}$K atoms from a cold atomic beam.
Afterwards, the atoms are further cooled to sub-Doppler temperatures in a molasses stage and optically pumped into the low-field seeking states $\ket{F=2, m_F=2}$ in order to load them into a harmonic magnetic trap, provided by the atom chip.
Here, the  $^{87}$Rb atoms are evaporatively cooled by removing the high-energetic atoms with a microwave knife and subsequent thermalization. 
Due to interspecies elastic collisions, the $^{41}$K atoms are cooled sympathetically until both species finally form a BEC \citep{modugno2001}. 
The magnetic traps containing the condensates are moved to their final position, where the atoms are magnetically collimated to achieve ultralow expansion velocities. Subsequently, the atoms are prepared in the magnetically insensitive state $\ket{F=2, m_F=0}$ by microwave pulses and released into free fall. 
With absorption imaging, the atoms can be spatially resolved using a CMOS camera (\textit{FLIR Grasshopper GS3-U3-23S6M}). The frame rate of the camera can be increased up to 1178$\,$Hz by reducing the region of interest of the sensor. Thus, it is possible to detect the different output ports of the atom interferometer encoded by their internal states by fast sequential imaging. Due to the fast loading of the MOT and high thermalization rates on the atom chip trap, the formation of BECs is reached for both species after a few seconds \citep{PiestDiss}, enabling many consecutive runs during the parabola flight.

\section{Description of Apparatus}
\label{sec:apparatus}
\begin{figure}
	\centering
	\def\svgwidth{0.9\columnwidth}
	\graphicspath{{}}	
	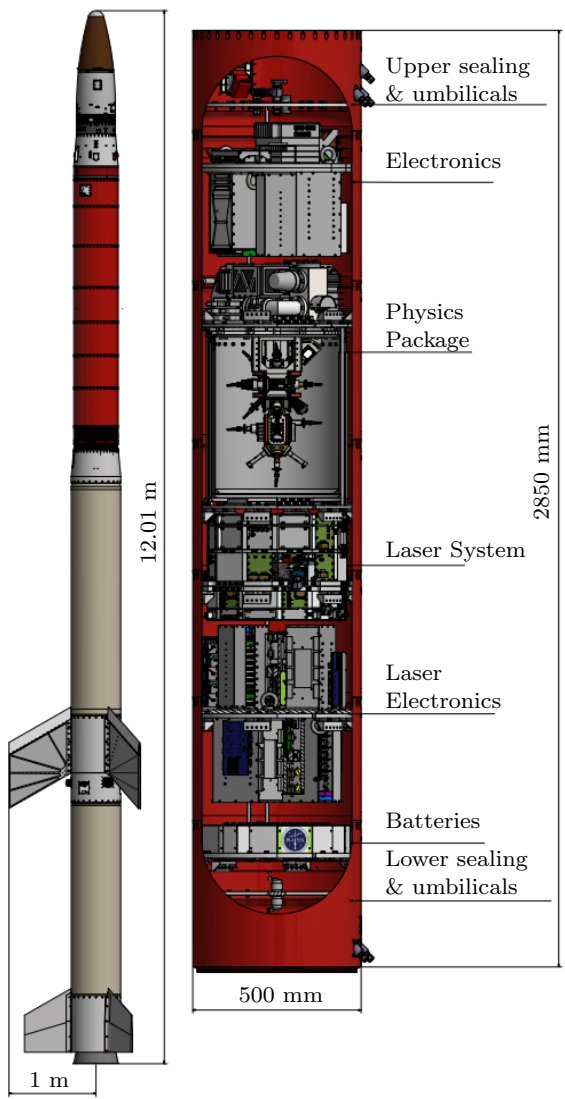
	\caption{
	The scientific payload (in red) on top of the two engines of a VSB-30 sounding rocket (left) and detailed view of the experimental payload (right). The payload hull is separated into seven RADAX segments. Image from \cite{ElsenDiss}.}
	\label{fig:pay}
\end{figure}

As shown in fig.\,\ref{fig:pay} (left) the scientific payload MAIUS-B (in red) is located on top of the two stages/engines of a VSB-30 sounding rocket. 
It offers a microgravity platform with residual accelerations in the 10$^{-5}$ to 10$^{-6}\,$g range \citep{Stamm2013} for about 325\,s during a parabola flight.

In the following section, the scientific payload with its subsystems is presented in detail. 

\subsection{Payload Overview}
\label{sec:overview}
In fig.\,\ref{fig:pay} (right) a detailled view of the scientific payload is shown.
It is separated into five subsystems which constitute a full atom interferometer apparatus using BEC mixtures. The subsystems provide the necessary laser frequencies,  electric currents for the manipulation of atoms as well as a thorough monitoring and housekeeping of the overall system. 
The subsystems are (from top to bottom): Electronics (EL), physics package (PP), laser system (LS), laser electronics (LE) and batteries (BA).

They are arranged in this order to place the center of gravity as close as possible (57$\,$mm, according to CAD model) to the position of the atoms.
This minimizes any effects of residual rotations of the payload on the outcome of the atom interferometer. These effects may enter via a shift of the free ensemble position during the sequence or a differential phase shift of the interferometer arms due to the Coriolis force \citep{hogan2008,Lenef1997}.

The rocket has a total length of \SI{12}{\meter} including \SI{2.8}{\meter} length of the scientific payload.

Since the experimental time in microgravity depends on the payload mass, it is essential to reduce it as much as possible. 
The mass budget presented in table \ref{tab:mass} with a total of \SI{335.3}{\kilogram} was obtained after a thorough optimization of the scientific payload design.
\begin{table}[h]
	\centering
	\caption{Mass budget of the MAIUS-B payload}
	\label{tab:mass}
	\begin{tabular}{p{0.2\textwidth}p{0.1\textwidth}p{0.1\textwidth}}
		\hline
		\noalign{\smallskip}
		System & Mass  \\
		& in \si{\kilogram}  \\
		\noalign{\smallskip}\hline \noalign{\smallskip}
		Electronics & \num{31.0}    \\
		Physics Package & \num{73.8}    \\
		Laser System & \num{53.5}    \\
		Laser Electronics & \num{35.2}  \\
		Batteries & \num{17.9}    \\
		Cables \& Cooling cycle & \num{35.7}   \\
		Sealing & \num{8.1}    \\
		Hull segments & \num{80.1}    \\
		\noalign{\smallskip}\hline \noalign{\smallskip}
		SUM & \num{335.3} \\
		\noalign{\smallskip}\hline
	\end{tabular}
\end{table}
It ensures a safe landing with the parachute system.
Additionally, the manacle-ring, which connects the scientific payload to the rocket motor of the sounding rocket, withstands the expected loads during launch. 

Each subsystem is suspended to the hull segments using six circularly arranged aluminum brackets. The brackets are equipped with two passive rubber vibration dampers and a safety bolt which limits the stretch of the dampers 
to prevent them from damage. Additionally, the bolts also serve as a safety measure in case of damper failure. Each suspension has to withstand static loads during ascent and re-entry of up to \SI{25}{\g} and up to \SI{100}{\g} shock at touchdown. FEM simulations show a safety factor per suspension bracket of more than 1.4.

\subsection{Physics Package}
\label{sec:pp}

\begin{figure*}
	\def\svgwidth{1.0\linewidth}
	\graphicspath{{}}	
	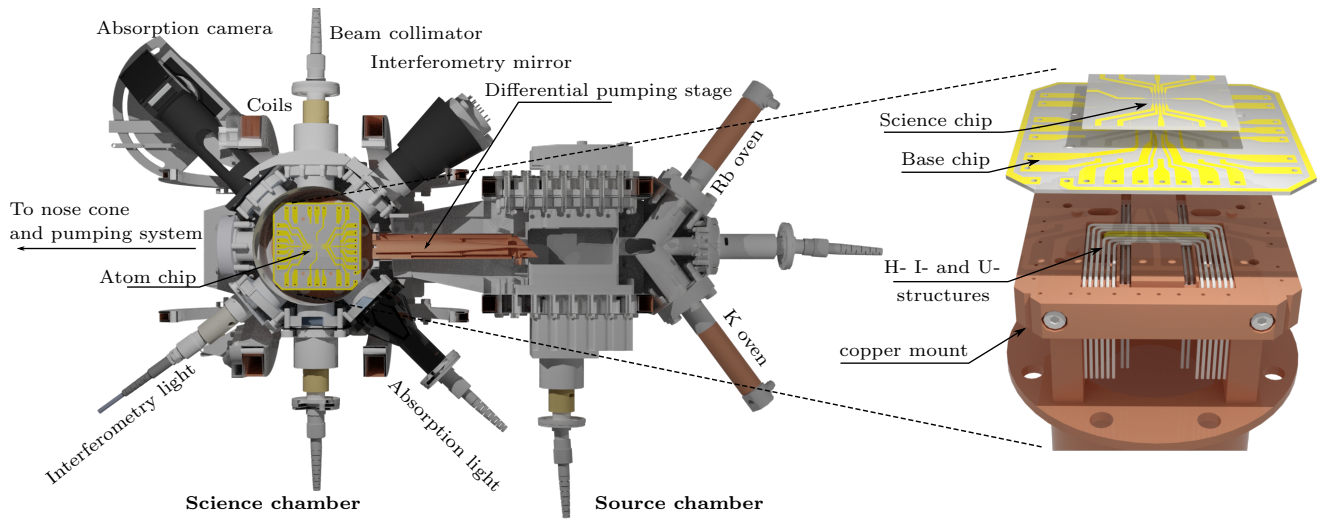
	\caption{
	Model of the experimental chamber (left) divided into science- and source chamber. Detailed view of the three-layer atom chip (right) }
	\label{fig:pp} 
\end{figure*}

The PP is divided into two main sections, the pumping system to maintain the vacuum and the experimental chamber with its external components where the experiments are performed.

The pumping system includes an ion getter pump (IGP), a titanium sublimation pump (TSP) and a pressure sensor. 
It is seperated from the experimental chamber to protect it from magnetic stray fields originating from the IGP. 
To further reduce magnetic fields affecting the atoms, the experimental chamber is surrounded by a three-layer magnetic shield of Mu-metal with a measured shielding factor of 10000 and 2000 in lateral and longitudinal direction, respectively \citep{kubelka2016}.

The experimental chamber inside the magnetic shield without the pumping system is shown in fig. \ref{fig:pp} (left). 
To separate the relatively high partial pressure area (source chamber) for the creation of a cold atomic beam from the science chamber with a pressure in the $10^{-11}$ mbar regime, a two chamber design has been chosen. The two chambers are connected via a differential pumping stage (DPS) which maintains their pressure difference.
Thus, the lifetime of the atomic ensembles in the science chamber can be increased due to less collisions with the background gas while maintaining a high atom flux from the source chamber.

As a reservoir, two ovens containing \SI{1}{\gram} of Rb and K respectively are connected by a flange to the source chamber. The K oven is kept at a temperature of \SI{46}{\celsius} while the Rb oven is left unheated at room temperature. 
In the source chamber the K and Rb atoms are confined in a 2D-MOT which directs them through the DPS into the center of the science chamber. 

To create high magnetic field gradients for atom trapping while keeping the switching times and current consumption low, an atom chip is mounted in the center of the science chamber, as shown in fig. \ref{fig:pp} (right). 
It consists of three layers: First, copper wires with a diameter of 1$\,$mm in U-, I- and H-shapes are mounted onto the holder which can be used to generate high volume quadrupole field configurations for the MOT or harmonic magnetic traps.
Above, a base chip with an electroplated z-shaped gold structure as well as two U-shaped loops for applying radio- or microwaves frequencies is glued. 
On top, a science chip with a 50$\,\mathrm{\mu m}$ wide z-shaped gold structure is closest to the atoms and allows for DC currents up to 2$\,$A to create magnetic traps with high trapping frequencies. A mirror coating on top of the science chip allows for light configurations addressing the atoms from all directions. 
With the combination of these structures high volume as well as high frequency traps can be realized. 
Additional coils around both chambers produce offset fields in the science chamber and a quadrupole field in race-track configuration in the source chamber. 

The light for cooling and manipulating the atoms is connected to the vacuum chamber via optical fibers and external optical setups to collimate and adjust the beams in diameter and shape.

Two perpendicular detection systems are used, one based on absorption imaging and the other one using the fluorescence of the atoms. 

For matter wave interferometry, a collimator (\textit{Schäf\-ter + Kirch\-hoff, 60FC-4-M15-37}) is attached to the vacuum chamber and retroreflected on the opposite side while passing through a quarter-wave plate twice.
The vibrations of the interferometry mirror are monitored using an accelerometer which is mounted rigidly behind the mirror.

\subsection{Laser System} \label{sec:las}
\begin{figure}
	\centering
	\def\svgwidth{0.9\columnwidth}
	\graphicspath{{}}	
\begingroup%
  \makeatletter%
  \providecommand\color[2][]{%
    \errmessage{(Inkscape) Color is used for the text in Inkscape, but the package 'color.sty' is not loaded}%
    \renewcommand\color[2][]{}%
  }%
  \providecommand\transparent[1]{%
    \errmessage{(Inkscape) Transparency is used (non-zero) for the text in Inkscape, but the package 'transparent.sty' is not loaded}%
    \renewcommand\transparent[1]{}%
  }%
  \providecommand\rotatebox[2]{#2}%
  \newcommand*\fsize{\dimexpr\f@size pt\relax}%
  \newcommand*\lineheight[1]{\fontsize{\fsize}{#1\fsize}\selectfont}%
  \ifx\svgwidth\undefined%
    \setlength{\unitlength}{512.15709776bp}%
    \ifx\svgscale\undefined%
      \relax%
    \else%
      \setlength{\unitlength}{\unitlength * \real{\svgscale}}%
    \fi%
  \else%
    \setlength{\unitlength}{\svgwidth}%
  \fi%
  \global\let\svgwidth\undefined%
  \global\let\svgscale\undefined%
  \makeatother%
  \begin{picture}(1,0.80221622)%
    \lineheight{1}%
    \setlength\tabcolsep{0pt}%
    \put(0,0){\includegraphics[width=\unitlength,page=1]{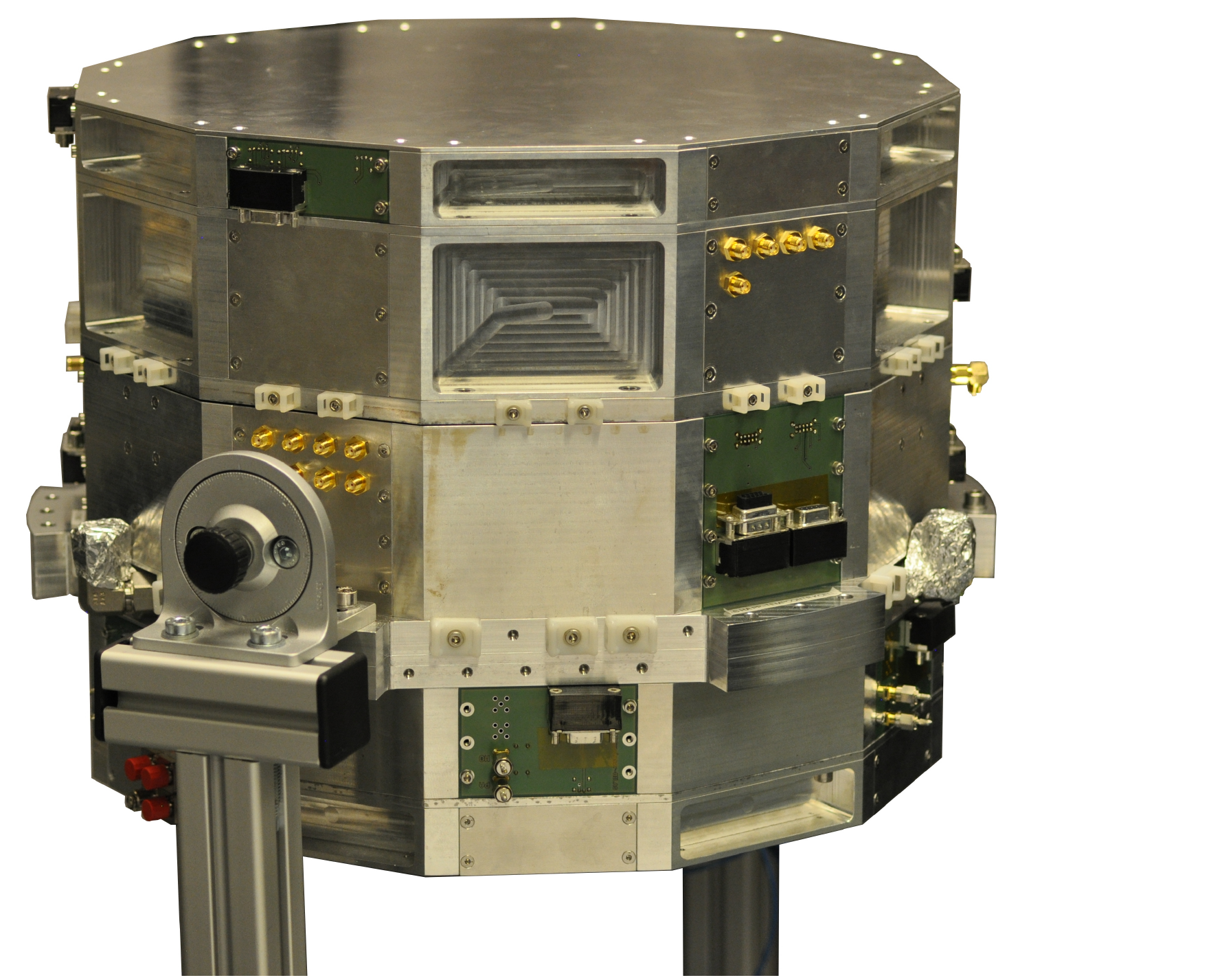}}%
    \put(0.92306971,0.17408566){\color[rgb]{0,0,0}\makebox(0,0)[lt]{\lineheight{0}\smash{\begin{tabular}[t]{l}f\end{tabular}}}}%
    \put(0.92306971,0.26872396){\color[rgb]{0,0,0}\makebox(0,0)[lt]{\lineheight{0}\smash{\begin{tabular}[t]{l}e\end{tabular}}}}%
    \put(0.92306971,0.36583724){\color[rgb]{0,0,0}\makebox(0,0)[lt]{\lineheight{0}\smash{\begin{tabular}[t]{l}d\end{tabular}}}}%
    \put(0.92306971,0.44839343){\color[rgb]{0,0,0}\makebox(0,0)[lt]{\lineheight{0}\smash{\begin{tabular}[t]{l}c\end{tabular}}}}%
    \put(0.92306971,0.61987878){\color[rgb]{0,0,0}\makebox(0,0)[lt]{\lineheight{0}\smash{\begin{tabular}[t]{l}b\end{tabular}}}}%
    \put(0.92306971,0.71699283){\color[rgb]{0,0,0}\makebox(0,0)[lt]{\lineheight{0}\smash{\begin{tabular}[t]{l}a\end{tabular}}}}%
    \put(0,0){\includegraphics[width=\unitlength,page=2]{Fig3.pdf}}%
  \end{picture}%
\endgroup%

	\caption{Photograph of the laser system. a) Distribution-module, b) Zerodur-module, c) reference-module, d) heatsink, e) laser-module, f) beat-module}
	\label{fig:123}
\end{figure}

The laser system provides the light fields at \SI{780}{\nano\meter} and \SI{767}{\nano\meter} for the experiments with Rb and K, respectively. 
This light enables laser cooling and optical pumping of the atoms as well as atom interferometry and imaging of the BECs. 
Additionally, light for an optional optical dipole trap at \SI{1064}{\nano\meter} is provided.
The system is driven and controlled by the laser electronics and delivers light via eleven polarization maintaining (PM) fibers directly to the optical ports of the collimators at the PP.
A photograph of the laser system is shown in fig.\,\ref{fig:123}.
It is assembled onto a \SI{30}{\milli\meter} thick baseplate which also serves as the heatsink.
It consists of five functional modules named (from bottom to top): beat-module, laser-module, reference-module, Zerodur-module and distribution-module.
The laser light originates from ten micro-integrated lasers which are mounted onto both sides of the heatsink.
Two of the ten laser modules are distributed feedback (DFB) lasers at \SI{780}{\nano\meter} and \SI{767}{\nano\meter}, which serve as reference lasers.
The other eight laser modules are based on a master-oscillator-power-amplifier-configuration (MOPA) concept, where an extended cavity diode laser (ECDL) serves as the master oscillator and a semiconductor optical amplifier provides the power boost.
The ECDL-MOPAs provide more than \SI{350}{\milli\watt} optical power ex fiber of spectral narrow light at \SI{767}{\nano\meter}, \SI{780}{\nano\meter}, and \SI{1064}{\nano\meter} \citep{Wicht2017s, Kuerbis2020} that is sent to the PP.

One ECDL-MOPA laser emits at \SI{1064}{\nano\meter} and provides light for the optical dipole trap. 
This laser is mounted in the reference module and is followed by a fiber pigtailed optical isolator mounted in the Zerodur-module, a fiber pigtailed acousto-optic modulator (AOM) for pulse shaping and a fiber switch to block residual light.
The two DFB reference lasers are mounted onto the top side of the heatsink in the reference module. 
Each reference laser is frequency stabilized to the most prominent spectroscopy transitions of $^{85}$Rb and $^{39}$K, respectively, by means of modulation transfer spectroscopy (MTS) \citep{shirley1982}. 
The MTS setups are implemented in one spectroscopy module per species, which are based on a robust Zerodur assembly technology \citep{Mihm2019, Duncker2014}. 
Similar frequency references were previously used on sounding rocket missions \citep{Lezius2016, Dinkelaker2017, Schkolnik2016}. 

Three ECDL-MOPAs emitting light at \SI{780}{\nano\meter} and four ECDL-MOPAs at \SI{767}{\nano\meter} are mounted in the laser-module onto the bottom side of the heatsink.
These seven ECDL-MOPA lasers are frequency stabilized in the beat-module with respect to the reference lasers with a dynamic frequency offset.
The main optical output power from the ECDL-MOPA modules operating at \SI{780}{\nano\meter} and \SI{767}{\nano\meter} is fiber coupled to five highly compact and robust optical benches.
These consist of Zerodur-made baseplates with glued-on optics for low-loss beam manipulation in free space, including superposition of light of different wavelengths in the same polarization state with dichroic mirrors and laser pulse shaping with AOMs.
The Zerodur-benches also feature mechanical shutters to block residual light that might harm the experiment when no light is required.
From the Zerodur-benches, light is guided via PM optical fibers into the distribution module. 
Here, the light fields for laser cooling of both atomic species are finally overlapped and split in a fiber beam splitter array to provide the required beam balance. 
The light-grid pulses for Raman double-diffraction interferometry of Rb and K are also overlapped in a fiber beam splitter, before passing a fiber pigtailed AOM for common pulse shaping and a fiber switch for blocking residual light.
From the distribution module, the light is finally guided to the PP via eleven PM optical fibers, each one connected to its respective collimator.
Temperature sensitive components, such as ECDL-MOPAs or fiber beam splitters, are equipped with individual temperature control provided by the laser electronics.
Multiple photodiodes and negative temperature coefficient (NTC) sensors allow for monitoring of the laser system's condition and troubleshooting.
The five functional modules are connected via hinges to allow opening of the highly integrated system for maintenance if necessary. All fiber connections are spliced to minimize losses.

\subsection{Electronics System}
\label{sec:el}
The electronic system consists of three subsystems, called batteries (BA), the laser electronics (LE) and general electronics (EL).
They are all mounted on base plates which also serve as heatsinks within the TCS.

The grounding of all subsystems is defined by the BA. However, a second galvanically isolated circuit is defined by a set of batteries sitting in the EL supplying the chip- and coil current drivers. This minimizes any effects of ground shifts of the overall system to the atoms.
Furthermore, for reasons of electromagnetic compatibility the use of switching regulators has been avoided throughout the whole payload whenever possible. 

In the following the three subsystems are discussed in more detail.

\subsubsection{Batteries}

The BA provides 24 outputs which distribute the voltages throughout the scientific payload. 
Each of these ports provide 6.6\,\si{V} and 20\,\si{V} and are controlled separately via the software. Furthermore, output currents and voltages are monitored for all ports.
The power is delivered by eight LiFePO$_4$ 15\,\si{Ah} batteries in two-cell and six-cell configuration. This leads to a possible autonomous operation time of \SI{40}{\minute} of the overall system, which is sufficient for the parabola flight with additional temporal margin. The batteries can be charged with up to 20\,\si{A}. 
To guarantee a safe operation, their charging and output currents are protected with 30\,\si{A} fuses and the individual outputs to the experiment with 10\,\si{A} fuses. 
Furthermore, the temperature within the module is monitored. 
In case of overheating, the charging current is additionally regulated by the firmware.

\subsubsection{Laser electronics}
The LE delivers currents for lasers, generates the control loops for frequency stabilization and monitors laser frequencies.
Additionally it synthesizes and amplifies radio frequencies (RF) to drive the AOMs and it contains a temperature control system to stabilize the temperature of the lasers and other sensitive optical components. 
It consists of two similar submodules mounted on opposite sides of the heatsink, one for the Rb and one for the K section of the lasersystem. 

To allow for a modular assembly of scientific electronics, stackable cards similar to the PC/104 format form basic building blocks of the MAIUS-B electronic system.
Each side of the LE contains two stacks formed by several cards. One of them contains current drivers for the power amplifier lasers, the other one current drivers for the diode lasers and their respective frequency controllers. 
The laser current drivers are designed to provide a low-noise current output up to 2\,\si{A} in order to operate ECDL-MOPA systems or DFB lasers.
The frequency controllers offer frequency stabilization of the reference lasers, control of frequency offsets between lasers and also allow for a phase lock loop (PLL) between two lasers. 
The latter is required to perform matter wave interferometry with Raman double-diffraction light pulses.

\subsubsection{General electronics}
\label{sec:electronics}
The EL supplies currents for the coils and for the atom chip structures of the PP. The chip and coil current drivers deliver up to 10\,\si{A} and are configured together in one stack. 
To minimize heating and losses of the magnetically trapped atoms \citep{Folman2002}, they feature a low-noise current source.
A second stack in this module houses the RF electronics and further auxiliary electronics. It generates microwave and RF signals needed for the evaporation and state preparation of the atoms. 
A dedicated atom chip protection module contains fuses and switching capabilities and additionally allows to monitor the currents delivered to the chip structures.
The EL houses the main computer which stores the predefined experimental sequences. Furthermore, it is responsible for housekeeping and telemetry data processing.

\subsubsection{Communication system}
Communication between the different electronic stacks is based on a network made of polymer optical fibers which has shown to have a number of advantages compared to a more conventional Ethernet-based approach \citep{oberschulte2021fpga}.
Besides the reduction of cable routing complexity and weight when using only a single wire, the use of plastic fibers also prevents ground loops between the connecting electronic components.
The used protocol combines clock recovery, in-band signaling, data transfer and prioritized trigger delivery over a single fiber.
The raw signal at \SI{150}{\mega\hertz} enables clock recovery and data transfer at \SI{50}{\mega\hertz}.
For in-band signaling and error detection 8b/10b line code is used as a transport-level protocol \citep{Widmer1983}.
On network layer four nibbles in each packet are used for the routing through the network from master nodes to endpoints with up to four hops. 
A single packet can be indefinitely long: Beginning and end are signaled by 8b/10b control words. Congestion is prevented by feedback control signals which temporarily stop the data flow in case of full buffers at the receiving end.
For the experiments simultaneous execution of programmed actions at the endpoints is important. 
Thus, a special, prioritized symbol is used as a global trigger; it will be transferred to all endpoints with a deterministic delay.

\subsection{Software}
\label{sec:soft}
The general architecture of the software inherit from the concepts initially developed for the MAIUS-1 mission \citep{Weps2018}. 
It consists of three main components: the experiment control software (ECS), the experiment design tools (EDT) and the ground control software (GCS) which are described below in detail.

The ECS has direct communication with the control electronics of the payload and its task is to execute the experiments as well as record data from the different subsystems.
Due to the large amount of hardware and domains involved, it has been chosen to use a model-driven approach. This is implemented for all hardware components and defines the possible control parameters for the experimentalist.
A minimal core system of the software provides the necessary interfaces to create new drivers and experiments. Their descriptions are given from the engineers and scientists using different domain-specific languages. For the drivers, the generated code is compiled with the rest of the software, while for the experiment descriptions they are interpreted in textual form.
Two main descriptions, called \textit{sequences} and \textit{graphs}, have been created to describe the experiments conducted in the flight. 
Sequences are abstract descriptions of single experiments.
Graphs are shaped as decision trees and use sequences as basic blocks. They allow a dynamic execution of the sequences depending on the outcome of previous experiments.

The EDT comprise the components used by the experimentalists in order to create or monitor experiments. For this, different graphical user interfaces allow to create new sequences and graphs in a convenient way. Also, machine learning algorithms are implemented to optimize the experimental parameters \citep{mloop}.

Finally, the GCS is the set of applications running on the ground station from where the mission will be controlled and monitored. 
It allows bi-directional communication with the ECS using telecommands. 
Thus, it is possible to upload sequences and visualize or store monitoring data during flight. 
However, the execution of the experiments is expected to be fully autonomous.

\subsection{Thermal Control System and Ground Support Equipment}
\label{sec:tcs}
The TCS of the MAIUS-B apparatus takes two operation modes into account, the laboratory and flight mode.
In laboratory operation the payload is actively cooled via the connected ground support equipment (GSE) and all components which produce heat within the payload are considered.
During ascent and re-entry aerodynamic drag adds a further heat source to the system. However, no active cooling is possible and the design of the TCS in flight mode relies solely on the thermal mass.

The estimated internally produced thermal loads are presented in Table 2.
\begin{table}[h]
	\centering
	\caption{Internal heat loads of the MAIUS-B apparatus}
	\label{tab:heat}
	\begin{tabular}{p{0.3\textwidth}p{0.1\textwidth}}
		\hline
		\noalign{\smallskip}
		System & Heat load \\
		 & in \si{\watt} \\
		\noalign{\smallskip}\hline \noalign{\smallskip}
		Electronics & \num{174.0}   \\
		Physics Package & \num{63.9}   \\
		Laser System & \num{100.0}    \\
		Laser Electronics & \num{340.9}  \\
		Batteries & \num{28.8}    \\
		\noalign{\smallskip}\hline \noalign{\smallskip}
		SUM & \num{707.6} \\
		\noalign{\smallskip}\hline
	\end{tabular}
\end{table}
Each subsystem except for the PP is equipped with an active liquid cooling with the respective baseplates acting as heatsinks. 
The design of the heatsinks for EL, LS, and LE include two base plates screwed together.
A meander is mounted inside these base plates and is connected to the intermodule coolant hoses.
The base plate of the PP is not equipped with liquid cooling.
Nevertheless, to support an efficient heat transfer off heat-producing amplifiers sitting on top of the PP, these are connected to an additional liquid cooled plate. 
With a comparatively low heat load of \SI{28.8}{\watt} the heatsink of the BA is made of a single-layer base plate with a meander mounted below.

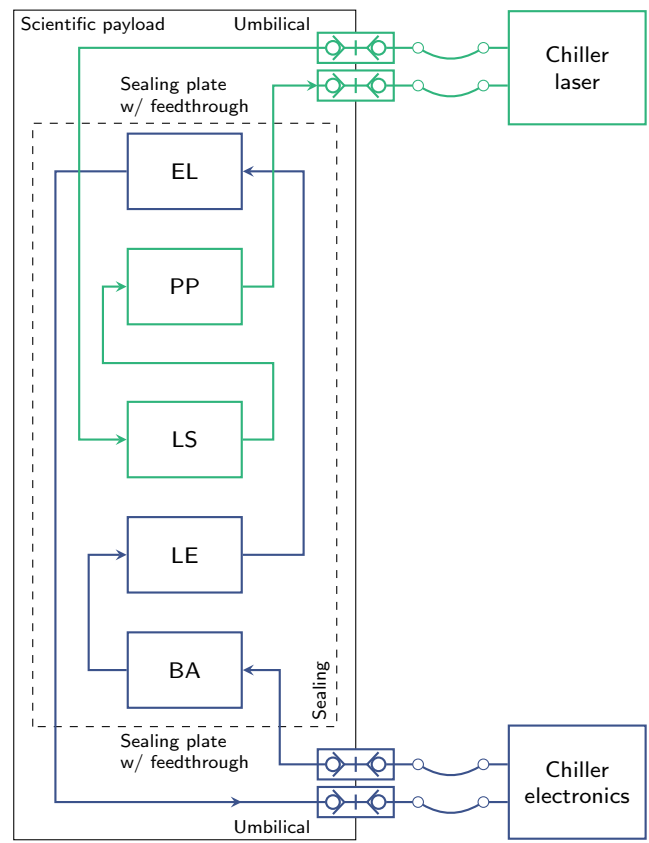
\begin{figure}[h]
	\centering
	\sffamily
	\scriptsize
	\begin{tikzpicture}[node distance=1cm]

    \node (payload) [below right, align=left] at (0, 11cm) {Scientific payload};
    \node (umbilical-down) [above left, align=right] at (4cm, 0) {Umbilical};
    \node (umbilical-up) [below left, align=right] at (4cm, 11) {Umbilical};
    \node (sealing-plate-down) [below, align=left] at (2.25cm, 1.5cm) {Sealing plate\\w/ feedthrough};
    \node (sealing-plate-up) [above, align=left] at (2.25cm, 9.5cm) {Sealing plate\\w/ feedthrough};
    \node (sealing) [above right, align=left, rotate=90] at (4.25cm, 1.5cm) {Sealing};

    \footnotesize

    \draw (0,0) rectangle (4.5cm, 11cm);
    \draw [dashed] (0.25cm, 1.5cm) rectangle (4.25cm, 9.5cm);
    \node (ba) [thick, rect-blue] at (2.25cm, 2.25cm) {BA};
    \node (le) [thick, rect-blue, above=0.5cm of ba] {LE};
    \node (ls) [thick, rect-green, above=0.5cm of le] {LS};
    \node (pp) [thick, rect-green, above=1cm of ls] {PP};
    \node (el) [thick, rect-blue, above=0.5cm of pp] {EL};

    \node (chiller-el) [rect-chiller, align=center, above right,
    					color=blue-viridis, text=black] at (6.5cm, 0) {Chiller\\electronics};
    \node (chiller-ls) [rect-chiller, align=center, below right,
    					color=green-viridis, text=black] at (6.5cm, 11) {Chiller\\laser};
    \pic [color=blue-viridis] (qd-ch-el-1) at (4.5, 0.5) {qd}; 
    \pic [color=blue-viridis] (qd-ch-el-2) at ($ (qd-ch-el-1-anchor) + (0,0.5) $) {qd}; 
    \pic [color=green-viridis] (qd-ch-ls-1) at (4.5, 10.5) {qd}; 
    \pic [color=green-viridis] (qd-ch-ls-2) at ($ (qd-ch-ls-1-anchor) + (0,-0.5) $) {qd};
    \pic [color=blue-viridis] (flex-ch-el-1) at ($ (qd-ch-el-1-right) + (0.75, 0) $) {flex};
    \pic [color=blue-viridis] (flex-ch-el-2) at ($ (qd-ch-el-2-right) + (0.75, 0) $) {flex};
    \pic [color=green-viridis] (flex-ch-ls-1) at ($ (qd-ch-ls-1-right) + (0.75, 0) $) {flex};
    \pic [color=green-viridis] (flex-ch-ls-2) at ($ (qd-ch-ls-2-right) + (0.75, 0) $) {flex};

    \draw[thick, color=blue-viridis] (qd-ch-el-1-right) -- (flex-ch-el-1-left);
    \draw[thick, color=blue-viridis] (qd-ch-el-2-right) -- (flex-ch-el-2-left);
    \draw[thick, color=green-viridis] (qd-ch-ls-1-right) -- (flex-ch-ls-1-left);
    \draw[thick, color=green-viridis] (qd-ch-ls-2-right) -- (flex-ch-ls-2-left);

    \draw[thick, color=blue-viridis] (flex-ch-el-1-right) -- (flex-ch-el-1-anchor -| chiller-el.west);
    \draw[thick, color=blue-viridis] (flex-ch-el-2-right) -- (flex-ch-el-2-anchor -| chiller-el.west);
    \draw[thick, color=green-viridis] (flex-ch-ls-1-right) -- (flex-ch-ls-1-anchor -| chiller-ls.west);
    \draw[thick, color=green-viridis] (flex-ch-ls-2-right) -- (flex-ch-ls-2-anchor -| chiller-ls.west);

    \draw[arrow1way, color=blue-viridis] (qd-ch-el-2-left) -- ++(-0.5, 0) -- ++(0, 1.25) -- (ba.east);
    \draw[arrow1way, color=blue-viridis] (ba.west) -- ++(-0.5, 0) |- (le.west);
    \draw[arrow1way, color=blue-viridis] (le.east) -- ++(0.8, 0) -- ($ (el.east) + (0.8, 0) $) -- (el.east);
    \draw[arrow1way, color=blue-viridis] (el.west) -- ++(-0.9375, 0) |- ($ (qd-ch-el-1-left) + (-1, 0) $);
    \draw[thick, color=blue-viridis] ($ (qd-ch-el-1-left) + (-1.1, 0) $) -- (qd-ch-el-1-left);

    \draw[arrow1way, color=green-viridis] (qd-ch-ls-1-left) -| ($ (ls.west) + (-0.625, 0) $) |- (ls.west);
    \draw[arrow1way, color=green-viridis] (ls.east) -- ++(0.4, 0) |- ($(pp.west)!0.5!(ls.west) + (-0.3125, 0) $) |- (pp.west);
    \draw[arrow1way, color=green-viridis] (pp.east) -- ++(0.4, 0) |- (qd-ch-ls-2-left);

	\end{tikzpicture}
	\caption{Schematic overview of the two cycle cooling concept within the TCS. The electronics cycle (in blue) feeds the subsystems BA, LE and EL and the laser cycle (in green) the LS and PP. Figure adapted from \cite{ElsenDiss}.}
	\label{fig:tcs}
\end{figure}

The TCS is divided into two cooling cycles which are connected via umbilicals to the GSE, as shown in fig.\,\ref{fig:tcs}. 

A \textit{Huber Unichiller 007} is used for the laser cycle and a \textit{Huber Unichiller 010} for the electronic subsystems. 
These chillers provide a maximum flow rate of $25\,$L/min and allow for pressure loss compensation of up to $2.5\,$bar. 
Since the GSE is placed on the launch pad within the exhaust gas of the rocket, the chiller is integrated into metal cases. 
These cases are made of aluminum and will be covered with a fire resistant pyroblanket to withstand the rough conditions inside the launch tower.
Due to the low ambient temperature at the launch site/landing area and to avoid damage caused by freezing of the remaining cooling liquid inside the payload, a thermo fluid (glycol/water mixture) is used. The mixture with \SI{44}{\percent} of glycol prevents freezing down to -\SI{30}{\degreeCelsius} and a mixture with \SI{52}{\percent} glycol down to -\SI{40}{\degreeCelsius}. This, however results in a higher viscosity of the cooling liquid as discussed in \ref{sec:flowtest}.

\subsection{Umbilicals and sealing}\label{sec:usc}
\begin{figure}[h]
	\centering
	\def\svgwidth{1.0\columnwidth}
	\graphicspath{{}}	
\begingroup%
  \makeatletter%
  \providecommand\color[2][]{%
    \errmessage{(Inkscape) Color is used for the text in Inkscape, but the package 'color.sty' is not loaded}%
    \renewcommand\color[2][]{}%
  }%
  \providecommand\transparent[1]{%
    \errmessage{(Inkscape) Transparency is used (non-zero) for the text in Inkscape, but the package 'transparent.sty' is not loaded}%
    \renewcommand\transparent[1]{}%
  }%
  \providecommand\rotatebox[2]{#2}%
  \newcommand*\fsize{\dimexpr\f@size pt\relax}%
  \newcommand*\lineheight[1]{\fontsize{\fsize}{#1\fsize}\selectfont}%
  \ifx\svgwidth\undefined%
    \setlength{\unitlength}{636.7500173bp}%
    \ifx\svgscale\undefined%
      \relax%
    \else%
      \setlength{\unitlength}{\unitlength * \real{\svgscale}}%
    \fi%
  \else%
    \setlength{\unitlength}{\svgwidth}%
  \fi%
  \global\let\svgwidth\undefined%
  \global\let\svgscale\undefined%
  \makeatother%
  \begin{picture}(1,0.67850951)%
    \lineheight{1}%
    \setlength\tabcolsep{0pt}%
    \put(0,0){\includegraphics[width=\unitlength,page=1]{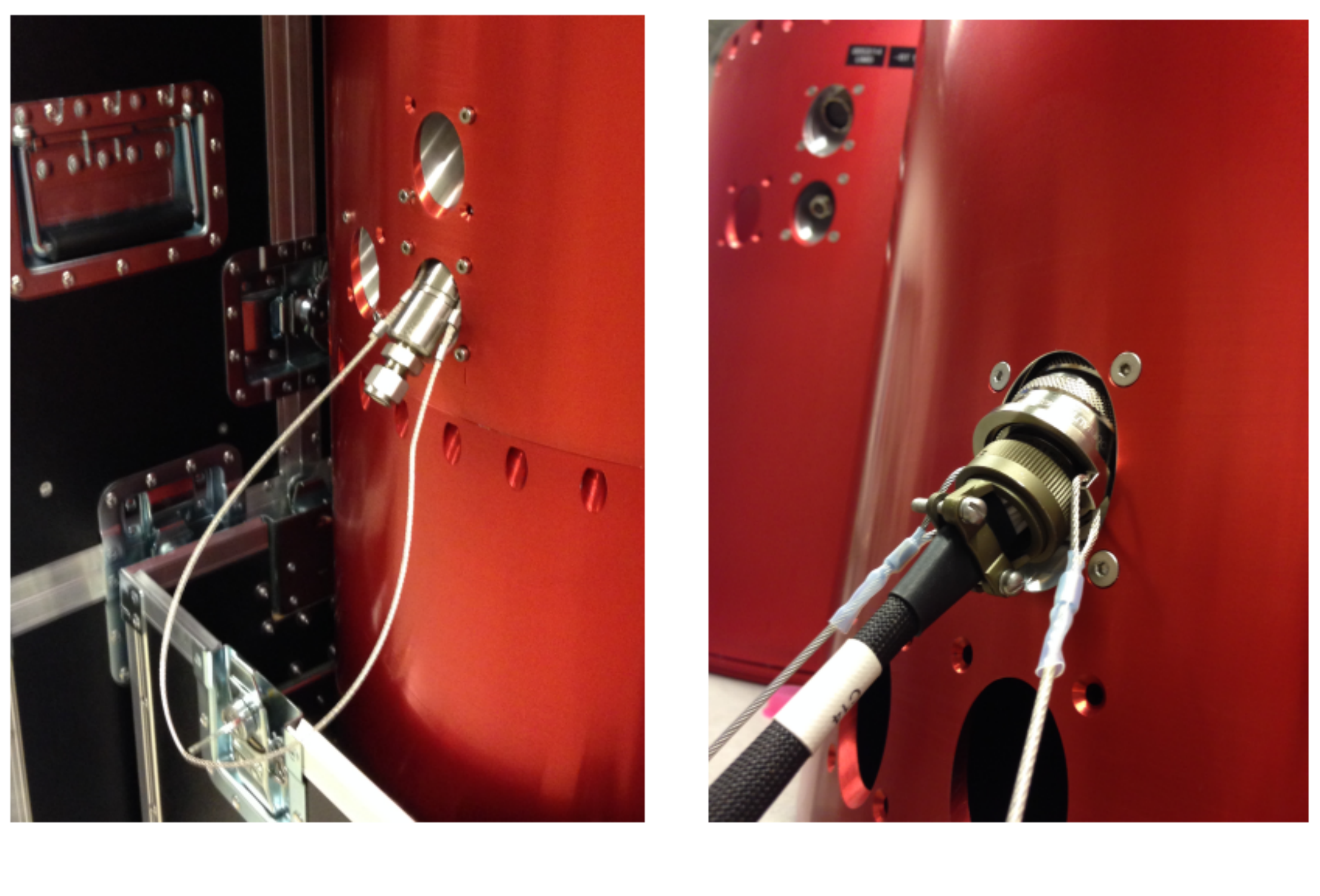}}%
    \put(0.08101972,0.01007622){\color[rgb]{0,0,0}\makebox(0,0)[lt]{\lineheight{1.25}\smash{\begin{tabular}[t]{l}Water umbilical\end{tabular}}}}%
    \put(0.59119976,0.01007622){\color[rgb]{0,0,0}\makebox(0,0)[lt]{\lineheight{1.25}\smash{\begin{tabular}[t]{l}Power umbilical\end{tabular}}}}%
  \end{picture}%
\endgroup%

	\caption{
	Coolant (left) and power umbilicals (right). The umbilicals for coolant liquid are equipped with a \textit{Swagelok} quick-coupling. The electrical umbilicals use push-and-pull quick connectors from \textit{Souriau}. Both are mounted at the upper and lower hull segments. Picture from \cite{GrosseDiss}.}
	\label{fig:umb}
\end{figure}

After the subsystems are integrated in flight configuration into the hull segments, the payload needs to be accessible for coolant, power and data connection. 
Several umbilicals are connected until lift-off to ensure active liquid cooling of the TCS, monitor and charge the batteries, enable data transfer, and allow an activation of the TSPs within the PP.
As shown in fig.\,\ref{fig:umb}, the quick connectors are equipped with a lanyard steel cable which will be attached to the launch pad.
This will unlock the plug automatically by the pull of the rocket as it lifts off during launch.

The payload is pressurized with artificial air to ensure a constant and reproducible environment for the LS.
To this end, it is sealed with sealing plates on the bottom and top hull segments using rubber O-rings. 
The seven hull segments (cf. fig. \ref{fig:pay}) are connected to each other by standardized RADAX flanges, also sealed with rubber O-rings.
To pressurize the payload, both sealing plates are equipped with a \textit{Swagelok} bulkhead fitting and ball tap as well as further feedthroughs for liquid cooling and power/data connection.
All power and data connections use hermetic military grade connectors. 
Cables and hoses from the umbilical to the GSE need to withstand the high temperatures during launch. Thus, these lines are equipped with a fire protection which is rated up to \SI{200}{\degreeCelsius}.

\section{Qualification Process}
\label{sec:Qualification}
The qualification process starts with pressuring tests of the hull segments to verify the integrity of the sealed payload compartment. Moreover the TCS is tested by measuring the flow rate of the cooling liquid and simulated via finite-element method (FEM) analysis. 
Furthermore, random frequency vibration tests on component and subsystem level are conducted.
Prior to flight campaign, the overall payload in launch configuration requires additional tests including a bench test, a spin balance test and further vibration tests.
In the following sections these tests are discussed in more depth.

\subsection{Payload pressuring test}
\label{sec:Ptest}
To measure the pressure loss over time, the hull segments including the sealing plates prepared in flight configuration are pressurized with artificial air (\SI{20}{\percent} oxygen and \SI{80}{\percent} nitrogen). 
To simulate the conditions during the actual flight, where the pressure difference between the outside (vacuum) and the inside of the payload (atmospheric pressure) will be around 1000 - \SI{1100}{\hecto\pascal}, the payload will be pressurized to an absolute pressure of \SI{2100}{\hecto\pascal}.
The pressure is monitored and logged for \SI{90}{\hour}. During the measurement, a pressure loss of \SI{40}{\hecto\pascal} has been detected. 
This shows that no significant leakages are expected before and during flight. 

\subsection{Flow rate measurement}
\label{sec:flowtest}

For the TCS and as an input for thermal simulations it is important to measure the flow rate of the cooling cycles.
In general, a higher viscosity of the cooling liquid leads to a reduction of the flow rate.
A too small flow rate could possibly cause problems in transporting heat out of the system.  The viscosity of glycol at room temperature is 21 times higher compared to water \citep{Tsierkezos98}. 
Therefore, two different glycol-water mixtures are tested (44\,\% and 52\,\% glycol) to measure the effect of the higher viscosity.
To measure the flow rate of the chiller without the cooling cycle as a reference, a test with a flow rate sensor and two short connection hoses (\SI{0.8}{\meter} each) is set up.
The test shows a flow rate of \SI{6.4}{\liter/\minute}. 

\begin{figure}[h]
	\includegraphics[width=0.9\columnwidth]{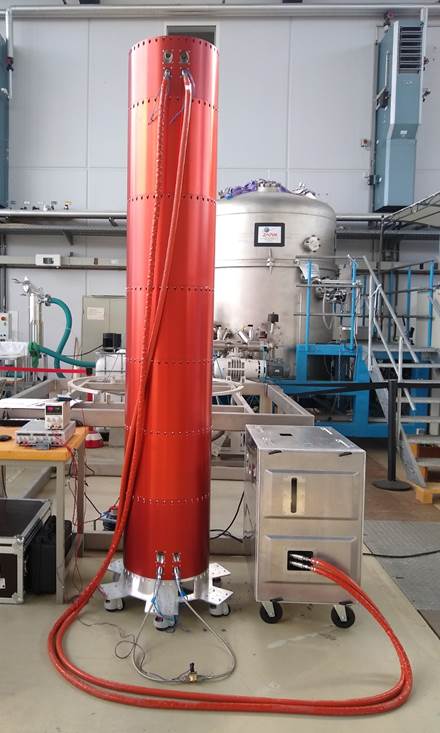}
	\caption{Flow rate test of the cooling cycles of MAIUS-B.
	The case which houses the chiller is placed right next to the scientific payload (red hull segments). The orange hoses are equipped with a fire protection to avoid damage during launch. Picture from \cite{ElsenDiss}.}
	\label{fig:flowrate}      
\end{figure}
To simulate a realistic setup of a single cooling cycle without the subsystems being included, the length of the cooling hoses within the hull is augmented to \SI{21}{\meter} including similar amount of fittings. 
Further, the inlet and outlet of the chiller are connected with the actual hoses as shown in fig. \ref{fig:flowrate}.
On the opposite side, both hoses are mounted to the upper umbilicals to simulate the most disadvantageous configuration. 
The flow rate sensor is connected with two \SI{0.8}{\meter} hoses directly to the lower coolant umbilicals.
As a result, flow rates of up to \SI{5.2}{\liter/\minute} (\SI{44}{\percent} glycol) and \SI{4.8}{\liter/\minute} (\SI{52}{\percent} glycol) are measured. 
This shows that the percentage of glycol has a direct impact on the flow rate. 
Nevertheless, FEM simulations based on these results show that the flow rate is sufficient to operate the TCS.

\subsection{Vibration Tests}
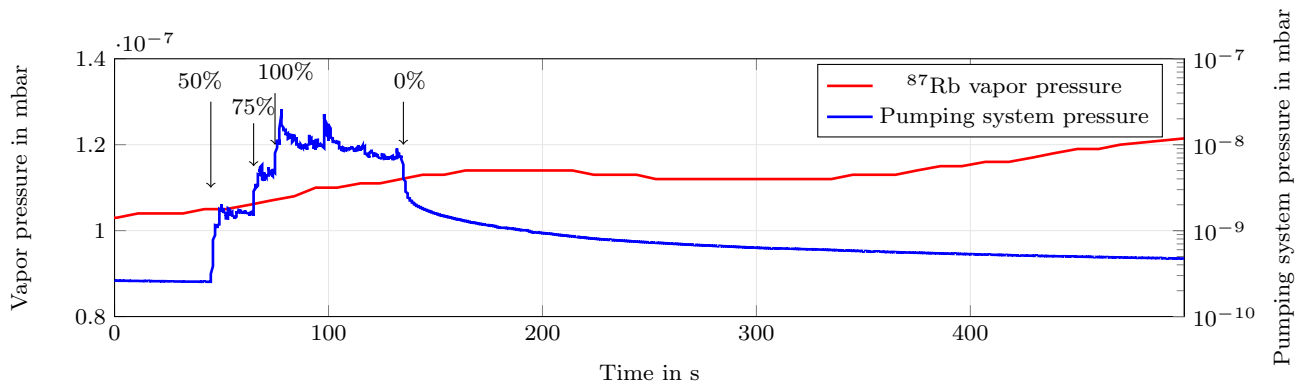
\begin{figure*}[t]
	\begin{tikzpicture}
		\begin{axis}[
			xlabel={Time in s},
			width=0.9\linewidth,
			height=5cm,
			ylabel={Vapor pressure in mbar},
			xmin=0, xmax=500, 
			ymin=0.8E-7, ymax=1.4E-7,
			ytick distance=0.2E-7,
			xtick={0,100,200,300,400},
			legend pos=south west,
			grid=both, major grid style = {line width=.2pt, draw = gray!20!}, minor grid style = {line width=.2pt, draw = gray!00!},		
			]
			\addplot[red,solid,line width=1.0pt]
			table[x=seconds since T0, y=partial87, col
			sep=comma,]{partialpressure_tag2.csv};
			\label{Partial pressure}
			\node[black] at (axis cs: 40,1.35E-7) {$50\%$};
			\node[black] at (axis cs: 65,1.29E-7) {$75\%$};
			\node[black] at (axis cs: 80,1.37E-7) {$100\%$};			
			\node[black] at (axis cs: 138,1.35E-7) {$0\%$};

			\draw [->](axis cs:45,1.3E-7) -- (axis cs:45,1.1E-7);
			\draw [->](axis cs:65,1.25E-7) -- (axis cs:65,1.15E-7);
			\draw [->](axis cs:75,1.32E-7) -- (axis cs:75,1.2E-7);
			\draw [->](axis cs:135,1.3E-7) -- (axis cs:135,1.2E-7);
		\end{axis}
	
		\begin{semilogyaxis}[
			width=0.9\linewidth,
			height=5cm,
			hide x axis,
			axis y line*=right,
			xmin=0, xmax=500, 
			ymin=1E-10, ymax=1E-7,
			ylabel={Pumping system pressure in mbar},
			ylabel near ticks
			]
			\addlegendimage{/pgfplots/refstyle=Partial pressure}\addlegendentry{$^{87}$Rb vapor pressure}	
			\addplot[blue,solid,line width=1.0pt]
			table[x=seconds since T0, y=Druck, col
			sep=comma,]{zShakerPressure_test1.csv};	
			\addlegendentry{Pumping system pressure}
		\end{semilogyaxis}
		
	\end{tikzpicture}
	
	\caption{Vacuum pressure in PP pumping system (blue curve) and $^{87}$Rb vapour pressure in the source chamber (red curve) prior, during and after vibrational test of the longitudinal axis. The rms-value of vibration loads is increased to the final value in three steps, as indicated by the arrows. Figure adapted from \cite{PiestDiss}.}
	\label{fig:vib_vacuum}
\end{figure*}

A sequential vibration qualification on component, subsystem and payload level throughout the technology life cycle is approached in this project. 
All vibration tests are conducted for a duration of \SI{60}{\second} for all three axes in a frequency range of $20-2000\,$Hz. 
Within component level, all critical parts are tested hard-mounted with a root mean square (RMS) value  of \SI{8.1}{\g} to minimize the risk of a failure during following subsystem or payload level tests. 
The subsystem level tests were performed in a cylindrical shaker adapter using the identical mounting as in the hull segments including a vibration isolation. The qualification on subsystem level is achieved with the vibration test profile given in table \ref{tab:vibprofile}. 
\begin{table}[h]
	\caption{Random vibration test profiles on subsystem level with power spectral density (PSD) and RMS values}
	\label{tab:vibprofile}
	\begin{tabular}{llll}
		\hline
		\noalign{\smallskip}
		\multicolumn{2}{c}{\textbf{Lateral}} & \multicolumn{2}{c}{\textbf{Longitudinal}}\\
		\noalign{\smallskip}\hline \noalign{\smallskip}
		Frequency & PSD & Frequency & PSD \\
		in \si{\hertz} & in \si{g\squared\per\hertz} & in \si{\hertz} & in \si{g\squared\per\hertz}\\
		\noalign{\smallskip}\hline \noalign{\smallskip}
		\num{20} - \num{400} & \num{0.002} & \num{20} - \num{900} & \num{0.002} \\
		\num{400} - \num{600} & \num{0.03} & \num{900} - \num{1600} & \num{0.01}\\
		\num{600} - \num{1300} & \num{0.002} & \num{1600} - \num{1800} & \num{0.06}\\
		\num{1300} - \num{2000} & \num{0.03} & \num{1800} - \num{2000} & \num{0.03} \\
		\noalign{\smallskip}\hline \noalign{\smallskip}
		RMS Value & \SI{5.41}{g} & RMS Value & \SI{5.18}{g} \\
		\noalign{\smallskip}\hline
	\end{tabular}
\end{table}The actual vibrational loads during flight are expected to be one-third of the qualification level \citep{GrosseDiss}.
Before and after each random test a resonance sine frequency sweep from \SI{20}{\hertz} to \SI{2000}{\hertz} with an amplitude of \SI{0.25}{g} and a sweep rate of \SI{2}{Oct/min} is performed. By comparing the measured frequency responses of the payload during the two resonance runs the test specimen are screened for structural flaws.\\
Above structural changes, further system specific parameters or functions have been monitored and checked for degradation or malfunction due to vibrational loads. For the laser system, parameters like optical powers along the various beampaths including coupling efficiencies and the alignment of optical components are of particular interest.\\
In case of the physics package, the vacuum pressure is a critical parameter which should be in a range capable for BEC production and atom interferometry experiments. Also the effect of the vibrations on the vapour pressure of $^{87}$Rb within the source chamber is probed. Results for the longitudinal vibration profile are shown in fig. \ref{fig:vib_vacuum}. The vapour pressure is measured spectroscopically by monitoring the absorbed light due to the D2-transition which is scanned by an external spectroscopy laser. Given the number of absorbing atoms and the temperature of the source chamber, the vapour pressure can be inferred by using the ideal gas law. The observed rise of the vapour pressure can not necessarily be attributed to the vibrations due to additional heat transfer from the vibration table and air condition to the experiment. However, it could be demonstrated that no significant increase, decrease or jumps of the vapour pressure are expected after rocket launch which could alter the atomic flux during flight experiments.\\
At the time we performed the vibration tests, the measured vacuum pressure was in the low $10^{-10}\,$mbar regime which is an order of magnitude above the current value. The expected lifetime of magnetically trapped Rb or K due to background gas collisions with hydrogen molecules is in the order of several seconds for a background pressure of $10^{-9}\,$mbar \citep{Folman2002}. Due to the much better vacuum conditions at present, it is expected that these results will be outperformed by future vibration tests.

\section{Summary and Outlook}
\label{sec:Summary}
A payload for parabolic flights on a VSB-30 sounding rocket is presented which is able to generate ultracold mixtures of $^{41}$K and $^{87}$Rb and perform matter wave interferometry in microgravity.
The apparatus consists of five modules which condense the functionality of a lab filling setup into the challenging constraints of a sounding rocket.
To allow an autonomous operation during the parabolic flight, the software executes a graph which determines the chronology of experimental sequences based on a decision tree. Possible self-optimization of the setup based on machine learning is implemented for ground operation.
To cope with the expected heat load of approximately $708\,$W, the TCS is designed to actively cool the experiment in laboratory operation and passively during flight. 
All subsystems are expected to withstand the static and dynamic loads and work within the operational requirements during launch and re-entry which has been demonstrated in dedicated vibration tests.

To test the functional interaction of the payload in its final configuration during flight, further qualification tests are going to be carried out:
In the bench test all payload systems including the scientific payload, the recovery system, the ignition unit, the attitude control system, and service module are brought together and are operated in flight like configuration.
Its aim is to screen for interferences between the payload systems, test the on-board software and to train the personnel in the procedures performed prior to and during flight.\\
The spin and balance test is a procedure to ensure the stability of the payload during ascent. Here, the payload is rotated with up to \SI{2}{\hertz} along its longitudinal axis to determine static and dynamic unbalances from the forces observed, which are compensated by installing counter weights. 
The maximal rotation rate during ascent of the rocket is expected to be \SI{2.7}{\hertz}, thus the mechanical stability of the payload components under flight-like conditions is also tested during spin balancing.\\
The findings of the planned experiments aboard the MAIUS-2/3 missions will advance the knowledge of the physics of ultracold quantum gases in microgravity. 
Further, a successful mission presents a technological milestone for possible future applications of cold quantum gases in orbital platforms including inertial sensing, space geodesy up to gravitational wave detection.

\begin{acknowledgements}
The QUANTUS IV - MAIUS project is a collaboration of Zentrum f\"ur angewandte Raumfahrttechnologie und Mikrogravitation Bremen, Leibniz Universit\"at Hannover, Humboldt-Universit\"at zu Berlin, Johannes Gutenberg-Universit\"at Mainz and  Fer\-di\-nand-Braun-Ins\-ti\-tut, Leibniz-Institut f\"ur H\"ochst\-fre\-quenz\-tech\-nik. It is supported by the German Space Agency DLR with funds provided by the Federal Ministry for economic affairs and climate action (BMWK) under grant number DLR 50WP 1431-1435.
We acknowledge support from Deut\-sches Zentrum f\"ur Luft- und Raumfahrt - Raumfahrtbetrieb, Oberpfaffenhofen, Deut\-sches Zentrum f\"ur Luft- und Raumfahrt - Simulations- und Softwaretechnik, Braunschweig.
Funded by the Deutsche Forschungsgemeinschaft (DFG, German Research Foundation) under Germany’s Excellence Strategy – EXC-2123 QuantumFrontiers – 390837967
\end{acknowledgements}

\section{Declarations}
\textbf{Funding:} This work is supported by the German Space Agency DLR with funds provided by the Federal Ministry for economic affairs and climate action (BMWK) under grant number DLR 50WP 1431-1435. We acknowledge support by the Deutsche Forschungsgemeinschaft (DFG, German Research Foundation) under Germany’s Excellence Strategy – EXC-2123 QuantumFrontiers – 390837967. \\ 
\textbf{Conflicts of interest/Competing interests:} The authors declare that they have no conflict of interest.\\
\textbf{Availability of data and material:} Corresponding contact Michael Elsen.\\
\textbf{Code availability:} Not applicable.\\
\textbf{Authors' contributions:} Michael Elsen and Baptist Piest wrote the manuscript, with contributions from all authors. Michael Elsen, Baptist Piest, Oliver Anton, Pawe\l{} Arciszewski, Wolfgang Bartosch, Dennis Becker, Jonas Böhm, Sören Boles, Klaus Döringshoff, Jens Grosse, Priyanka Guggilam, Ortwin Hellmig, Isabell Imwalle, Simon Kanthak, Christian Kürbis, Maike Diana Lachmann, Moritz Mihm, Alexandros Papakonstantinou, Christian Reichelt, Marvin Warner, Thijs Wendrich and André Wenzlawski designed, built and tested the apparatus. Fabian Adam, Pawe\l{} Arciszewski, Matthias Koch, Hauke Müntinga, Ayush Mani Nepal, Tim Obertschulte, Peter Ohr, Arnau Prat, Jan Sommer and Christian Spindeldreier developed and implemented the software. Claus Braxmaier, Holger Blume, Daniel Lüdtke, Achim Peters, Ernst Maria Rasel, Klaus Sengstock, Andreas Wicht and Patrick Windpassinger are the co-principal investigators of the project, and Jens Grosse is principal investigator.\\
\textbf{Ethics approval:} The authors confirm that the manu\-script is original and has not been published elsewhere or is it currently under consideration for publication elsewhere.\\
\textbf{Consent to participate:} Not applicable.\\
\textbf{Consent for publication:} All authors provided the consent for publication.\\
\bibliographystyle{spbasic}      
\bibliography{references}   

\end{document}